\documentclass[a4paper,11pt]{article}
\usepackage{pos}
\usepackage{bm}
\usepackage{microtype}
\usepackage{placeins}
\hypersetup{pdftitle={Hard Probes of QGP over the Decades},
 pdfauthor={Xin-Nian Wang},
 pdfsubject={Hard Probes 2026 proceedings},
 pdfkeywords={quark-gluon plasma, jet quenching, transport, medium response, diffusion wake}}

\newcommand{\dd}{\mathrm{d}}
\newcommand{\qhat}{\hat q}
\newcommand{\sNN}{\sqrt{s_{\mathrm{NN}}}}
\newcommand{\pT}{p_{\mathrm{T}}}

\title{Hard Probes of QGP over the Decades}
\ShortTitle{Hard Probes of QGP over the Decades}
\author{Xin-Nian Wang}
\affiliation{Central China Center for Nuclear Theory, Key Laboratory of Quark and Lepton Physics (MOE) \& Institute of Particle Physics,\\
Central China Normal University, Wuhan 430079, China}
\emailAdd{xnwang@ccnu.edu.cn}

\abstract{In this opening talk, I follow the historical perspective of a recent review \cite{WangWiedemann} and look back at the development of hard probes of the quark--gluon plasma (QGP), from Bjorken's proposal of parton energy loss to quantitative studies of jet transport and jet-induced medium response.  I discuss coherent medium-induced gluon radiation and the experimental observation of jet quenching at RHIC and the LHC. Comparisons of calculations with hadron and jet spectra and correlations constrain the jet transport coefficient, while heavy flavor and quarkonia probe diffusion, screening and recombination. The energy and momentum lost by energetic partons also excite the QGP medium, leading to modifications of soft hadron distributions and jet substructure. Recent measurements of the diffusion wake provide further information on this response and open a new window for the study of QGP properties. Quantitative extraction of QGP properties in the future should combine analyses of hard and soft observables with a consistent description of parton showers, the evolving medium and its response, with controlled theoretical and experimental uncertainties.}

\FullConference{13th International Conference on Hard and Electromagnetic Probes of\\
High-Energy Nuclear Collisions (HardProbes2026)\\
21--26 June 2026\\
Vanderbilt University, Nashville, Tennessee, USA}

\begin{document}
\maketitle

\section{From QCD to hard probes of the quark--gluon plasma}

Hagedorn's statistical bootstrap picture suggested a limiting temperature for a hadronic description of the strong interaction, while asymptotic freedom in QCD provided the basis for a high-temperature phase with quark and gluon degrees of freedom. Hard processes probe the properties of quark--gluon plasma (QGP) in high-energy nuclear collisions that requires controlled descriptions of particle production and medium modification. I discuss these developments in this talk following the historic perspective of Ref.~\cite{WangWiedemann}.

Two preprints by Bjorken in the summer of 1982 introduced two complementary approaches to QGP studies: a boost-invariant description of the longitudinal expansion~\cite{BjorkenHydro}, and elastic energy loss of energetic partons propagating through the plasma~\cite{BjorkenLoss}. Hard partons are produced before the QGP formation and the initial production rates provide a reference for modifications along their paths through the medium. Nuclear modifications of the incoming parton distributions should be taken into account to understand the final medium modification of the hard probes.

Inclusive hadron production can be organized schematically as
\begin{equation}
 \dd\sigma^{AA\to h+X}
 =\sum_{a,b,c,d} f_{a/A}\otimes f_{b/A}\otimes
 \dd\hat\sigma_{ab\to cd}\otimes \tilde{D}_{c\to h},
 \label{eq:factorization}
\end{equation}
where $f_{a/A}$ are nuclear parton distributions, $\dd\hat\sigma$ is the short-distance partonic cross section, and $\tilde{D}$ the medium modified fragmentation function due to parton energy loss and radiated gluons. The convolutions are over the initial and final parton momentum, and an average over the nuclear geometry within a given centrality class is implied.

Soft hadrons, electromagnetic radiation and hard probes provide complementary constraints on collective evolution and transport properties of the QGP medium and energetic jets. A hydrodynamic background constrained by soft hadron data is therefore needed for quantitative hard-probe studies. The early Hard Probe Collaboration and subsequent JET and JETSCAPE efforts coordinated these theoretical and experimental studies~\cite{WangWiedemann}.

\section{Parton energy loss and the experimental discovery of jet quenching}

Bjorken's original proposal emphasized collisional energy loss of energetic partons in QGP. The importance of non-Abelian bremsstrahlung led to studies of induced gluon radiation in the Gyulassy--Wang model, the BDMPS--Zakharov treatment of multiple scattering, opacity expansions and the higher-twist approach~\cite{GW,BDMPS,Zakharov,GLV,Wiedemann,GuoWang}. This has also been studied in the weak-coupling thermal field theory with in-medium splitting rates~\cite{AMY}. These approaches describe different limits of scattering, coherence and medium evolution.

Both elastic and inelastic energy loss involve the jet transport coefficient that characterizes the transverse momentum broadening squared per unit path length,
\begin{equation}
 \qhat_R=\frac{\dd\langle k_\perp^2\rangle}{\dd L}
 =\int\dd^2\bm q_\perp\,q_\perp^2
 \frac{\dd\Gamma_R}{\dd^2\bm q_\perp},
 \label{eq:qhat}
\end{equation}
where $L$ is path length and $\dd\Gamma_R$ is the scattering probability per unit length for a parton in color representation $R$. The jet transport coefficient depends on the projectile energy and resolution prescription. The quark and gluon transport coefficients, $\qhat_F$ and $\qhat_A$, are proportional to their respective  quadratic color Casimir, so that $\qhat_A/\qhat_F=C_A/C_F=9/4$ at leading order in $\alpha_{\rm s}$.

Radiation amplitudes from scatterings within the gluon formation time interfere leading to the Landau--Pomeranchuk--Migdal (LPM) effect. For a soft gluon with energy $\omega$, the relative phase is $k_\perp^2t/(2\omega)$. Multiple soft scattering in a static medium then gives
\begin{equation}
 t_f\simeq\frac{2\omega}{k_\perp^2},\qquad
 k_\perp^2\simeq\qhat_A t_f,\qquad
 t_f\simeq\sqrt{\frac{2\omega}{\qhat_A}},\qquad
 \omega_c=\frac12\qhat_A L^2.
 \label{eq:formation}
\end{equation}
The corresponding radiative energy loss has the parametric dependence
\begin{equation}
 \Delta E_{\mathrm{rad}}\sim\alpha_s C_R\qhat_A L^2,
 \label{eq:loss}
\end{equation}
for a projectile with $E\gg\omega_c$ in a static medium. Expansion, finite energy and fluctuations change the path-length dependence in heavy-ion collisions. The energy loss for a reconstructed jet also depends on angular acceptance, since some leading-parton radiation remains inside the jet cone.

\begin{figure}[t]
 \centering
 \includegraphics[width=\textwidth]{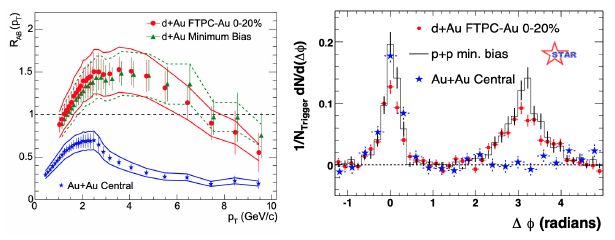}
 \caption{Nuclear modification factors for charged hadrons (left) and dihadron azimuthal correlations (right) in central Au+Au and d+Au collisions at RHIC. The strong suppression in Au+Au and its absence in the d+Au control measurements establish the final-state origin of the high $p_T$ hadron suppression. STAR results~\cite{STARdAu}, reproduced as assembled in Ref.~\cite{WangWiedemann}.}
 \label{fig:discovery}
\end{figure}

Parton energy loss was predicted to suppress leading hadron spectra~\cite{GWphen}. This suppression is measured through the nuclear modification factor,
\begin{equation}
 R_{AA}(\pT,y)=\frac{1}{\langle T_{AA}\rangle}
 \frac{\dd^2N_{AA}/\dd\pT\dd y}
 {\dd^2\sigma_{pp}/\dd\pT\dd y},\qquad
 \langle T_{AA}\rangle=\frac{\langle N_{\mathrm{coll}}\rangle}
 {\sigma^{\mathrm{inel}}_{pp}},
 \label{eq:raa}
\end{equation}
where $N_{AA}$ is the per-event hadron yield for a given centrality class. Binary collision scaling gives $R_{AA}=1$ in the absence of nuclear effects. Hadron suppression due to energy loss was first seen in experiments at RHIC as shown in Fig.~\ref{fig:discovery}. The suppression of high-$\pT$ hadrons and away-side correlations in central Au+Au, together with d+Au control measurements, established the final-state origin of the dominant suppression~\cite{PHENIX,STARdAu}. Measurements at the LHC extended the study to single inclusive hadrons, reconstructed single jets and dijets at higher energies~\cite{ATLASdijet,CMSdijet}. The dijet momentum imbalance is compensated by softer particles at large angles, indicating the need to include medium response for a full description of jet energy loss~\cite{CMSdijet}.

\section{Transport coefficients and complementary probes}

\subsection{From suppression to quantitative jet tomography}

Since parton energy loss depends on the initial jet production position and the space-time evolution of the QGP, an extraction of the transport coefficient requires a description of both. An energetic hadron trigger preferentially selects partons that lose less energy, while correlated hadrons and photon-tagged jets sample different distributions of production points, propagation lengths and initial energies. One can therefore constrain the jet transport coefficient more systematically by combining these measurements within a common description of the medium.

\begin{figure}[t]
 \centering
 \includegraphics[width=\textwidth]{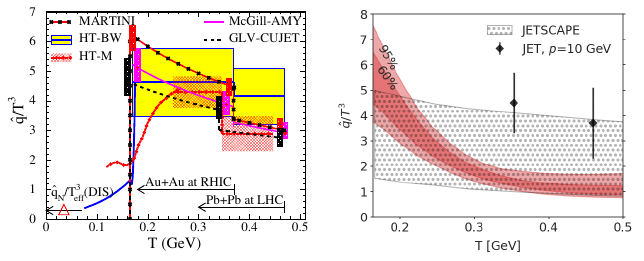}
 \caption{Scaled quark jet transport coefficient $\qhat_F/T^3$ extracted by the JET Collaboration with several energy-loss implementations (left)~\cite{JET}. The results of information-field Bayesian inference (right)~\cite{IFfull} are compared with the JET and JETSCAPE~\cite{JETSCAPE} results. The temperature dependence and uncertainty bands are obtained within the respective dynamical models and inference frameworks. Panels reproduced from Ref.~\cite{WangWiedemann}.}
 \label{fig:qhat}
\end{figure}

Shown in Fig.~\ref{fig:qhat} (left) are the JET Collaboration results from several energy-loss frameworks with a common hydrodynamic background constrained by bulk observables~\cite{JET}. For a 10~GeV quark at the center of the most central collisions and the initial time $\tau_0=0.6~\mathrm{fm}/c$, the extracted values are
\begin{equation}
 \frac{\qhat_F}{T_0^3}\simeq
 \begin{cases}
  4.6\pm1.2,&T_0=0.370~\mathrm{GeV}\quad\text{(RHIC)},\\
  3.7\pm1.4,&T_0=0.470~\mathrm{GeV}\quad\text{(LHC)}.
 \end{cases}
 \label{eq:jetvalues}
\end{equation}
 The model spread in Fig.~\ref{fig:qhat} illustrates part of the theoretical uncertainty.

In more recent studies, Bayesian analyses have been used to constrain the temperature and energy dependence of $\qhat$~\cite{JETSCAPE,IF,IFfull}. Most analyses use explicit parameterizations as priors that can correlate distant temperature regions, including those weakly constrained by data. The information-field approach reduces these long-range correlations in the prior~\cite{IF}. Different centralities and collision energies can then constrain different temperature regions with less dependence on a prescribed functional form. The combined single-hadron, dihadron and $\gamma$--hadron analysis in Fig.~\ref{fig:qhat} (right) favors a stronger temperature dependence of $\qhat/T^3$ than a constant-ratio ansatz~\cite{IFfull}. Future analyses should combine hard probes and  bulk hadron spectra and include uncertainties in energy loss, medium evolution, hadronization and experimental covariance.

\subsection{Heavy flavor and quarkonia}

The heavy-quark mass adds a scale for studying parton propagation and near-equilibrium diffusion in QGP. At moderate momenta, elastic scattering, collective flow and recombination with thermal quarks affect heavy-flavor spectra and anisotropies. For a nonrelativistic heavy quark, the equilibrium Einstein relation gives
\begin{equation}
 D_s=\frac{T}{M\eta_D}=\frac{2T^2}{\kappa}\qquad (p\to0),
 \label{eq:diffusion}
\end{equation}
where $D_s$ is the spatial diffusion coefficient, $M$ the heavy-quark mass and $\eta_D$ the drag coefficient. The momentum-diffusion coefficient is defined by the Langevin noise correlation $\langle\xi_i(t)\xi_j(t')\rangle=\kappa\delta_{ij}\delta(t-t')$~\cite{MooreTeaney}. This relation applies at low momentum.
At high momenta, induced gluon radiation and medium splitting become important. In this regime $\qhat$ in Eq.~\eqref{eq:qhat} characterizes the jet transport and is related to the high-energy limit of the momentum diffusion. Transport-model comparisons show that a common bulk evolution and controlled hadronization are needed to reduce the uncertainties in the extraction of the heavy quark diffusion coefficient~\cite{HFreview,CaoHF}.

Quarkonium suppression was proposed by Matsui and Satz as a probe of color screening in QGP~\cite{MatsuiSatz}. Later studies show, however, that the final yields depend on in-medium dissociation, regeneration, feed-down and the initial heavy-quark abundance~\cite{StatHadron}. Screening-length arguments alone therefore cannot determine the survival probability. Modern transport and open-quantum-system descriptions connect the in-medium spectral changes, dissociation and recombination to the measured yields. Together with light jets and open heavy flavor hadrons, these measurements probe different aspects of the QGP interaction.

\section{Jet-induced medium response and the diffusion wake}

Energy and momentum deposited in the medium by jets lead to medium response that modifies the QGP evolution. The coupled jet--medium evolution can be described by a relativistic viscous hydro with a source term in the soft sector and the linear Boltzmann equation for the hard partons,
\begin{equation}
 \partial_\mu T_{\mathrm{med}}^{\mu\nu}=\sum_i j^\nu(i)\theta(p_{\rm cut}^0-p_i\cdot u),
 \qquad p_i^\mu\partial_\mu f(p_i)={\cal C}[f(p_i)]\theta(p_i\cdot u-p_{\rm cut}^0),
 \label{eq:source}
\end{equation}
where $u^\mu$ is the local fluid velocity and $j^\nu(i)$ is the four-momentum density deposited into the medium by a parton with momentum $p_i^\nu$. These equations are the bases for the CoLBT-hydro model \cite{CoLBT}  which couples the linear Boltzmann transport (LBT) model \cite{LBT1,LBT2} to the CLVisc hydrodynamic model \cite{CLVsic} with concurrent evolution of both the hard and soft sectors of the QGP. The deposited energy and momentum by jets act as sources for the evolution of the expanding medium while the hard partons evolve further in the medium that is modified by the sources. Effects of medium response were also considered in the HYBRID model \cite{Hybrid}.

\begin{figure}[t]
 \centering
 \includegraphics[width=\textwidth]{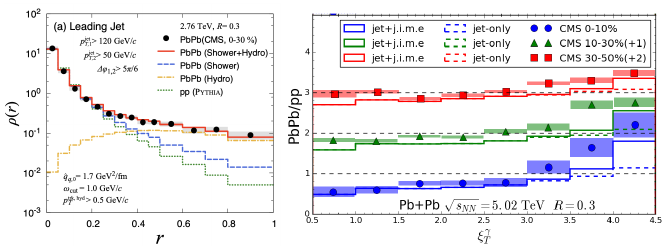}
 \caption{Left: jet shape $\rho(r)$, with $r=\sqrt{(\Delta\eta)^2+(\Delta\phi)^2}$, from a jet--fluid calculation~\cite{Tachibana}, compared with CMS data~\cite{CMSshape}. Right: Pb+Pb/pp ratio of photon-tagged fragmentation distributions versus $\xi_T^\gamma=\ln[-(p_T^\gamma)^2/(\bm p_T^h\cdot\bm p_T^\gamma)]$, calculated with and without jet-induced medium excitation~\cite{ChenResponse}, compared with CMS data~\cite{CMSfrag}. The 10--30\% and 30--50\% curves are shifted by $+1$ and $+2$. Panels reproduced from Ref.~\cite{WangWiedemann}.}
 \label{fig:response}
\end{figure}

The deposited energy-momentum by a supersonic jet excites a propagating sound with a Mach structure and a diffusion wake. Such a jet-induced medium response enhances soft hadron production at large angles and modifies jet shapes and fragmentation functions ~\cite{Tachibana,ChenResponse}, as shown in Fig.~\ref{fig:response}. It also changes the reconstructed jet energy for different jet radii. Since medium-induced gluon radiation also contributes to soft hadrons in the same phase space, it is difficult to isolate the effect of hydrodynamic response from measured hadron spectra associated with a hard jet.

A recent LBT calculation~\cite{ThermalJets} finds that the recoil medium partons can cluster into low-$\pT$, large-radius jets and enhance the yield of recoil jets at large angles. This interpretation in terms of thermal recoil jets predicts characteristic jet fragmentation functions and the jet shape which remain to be tested in future experimental measurements.

The diffusion wake, on the other hand, can lead to a depletion of soft hadrons in the azimuthal direction opposite to the propagating jet. CoLBT-hydro calculations predict a localized valley in the two-dimensional rapidity--azimuth correlation on top of a ridge from the underlying multiple parton interaction (MPI)~\cite{YangWake}. This structure is a unique signal of the diffusion wake but difficult to measure because the small signal-to-background ratio is only a fraction of a percent. In dijets, the wake of one jet normally overlaps the enhanced yield from the other. By separating the jets and the corresponding wake contributions in rapidity, one can construct a background-free observable of the diffusion wake. The proposed rapidity asymmetry which measures the difference in the rapidity structure of the diffusion wakes from dijets with different rapidity gaps, the underlying background is canceled~\cite{YangAsym,YangBG}. 

Let $C_X(\Delta\eta;g)$ denote the near-side yield per dijet per radian in $X=AA,pp$ collisions, for $|\Delta\phi|<\pi/2$ and dijet gap $g$. The corresponding medium-dependent rapidity asymmetry or double difference is~\cite{CMSwake}
\begin{equation}
 \Delta C(\Delta\eta;g)=
 \big[C_{AA}(\Delta\eta;g)-C_{AA}(\Delta\eta;g_0)\big]
 -\big[C_{pp}(\Delta\eta;g)-C_{pp}(\Delta\eta;g_0)\big],
 \label{eq:wake}
\end{equation}
where $g_0$ denotes $|\Delta\eta_{jj}|<0.5$. Events are reflected to enforce $\eta_{\mathrm{jet}_1}>\eta_{\mathrm{jet}_2}$.

\begin{figure}[t]
 \centering
 \includegraphics[width=.68\textwidth]{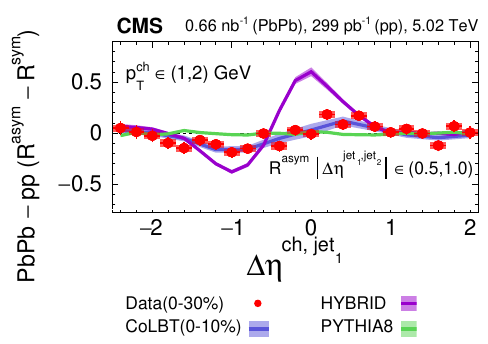}
 \caption{CMS yield double difference or rapidity asymmetry for $1<p_T^{\mathrm{ch}}<2$~GeV and $0.5<|\Delta\eta_{jj}|<1.0$ at $\sNN=5.02$~TeV~\cite{CMSwake} compared to CoLBT-hydro ~\cite{CoLBT} and HYBRID model \cite{Hybrid} calculations. The data are for 0--30\% centrality, while CoLBT-hydro is for 0--10\%. Adapted from the upper panel of Fig.~2 of Ref.~\cite{CMSwake}.}
 \label{fig:wake}
\end{figure}

As shown in Fig.~\ref{fig:wake} for $1<p_T^{\mathrm{ch}}<2$~GeV/$c$ in central Pb+Pb collisions at $\sNN=5.02$ TeV, the CMS Collaboration observes a diffusion wake depletion in gap intervals $0.5$--$1.0$. The signal exceeds five standard deviations in gap intervals $1.0$--$1.5$ and $1.5$--$2.0$~\cite{CMSwake}. This first observation of the jet-induced diffusion wake agrees quantitatively with CoLBT-hydro calculations (taking into account the different centrality selection of the data and calculation).

\FloatBarrier
\section{Outlook: toward precision studies of QGP}

Over four decades, the study of hard probes has progressed from a theoretical concept to identifying a dense QCD medium, constraining its microscopic interactions with energetic partons and collective response. Further progress requires consistent treatment of vacuum showers, nuclear initial-state effects, in-medium radiation, recoils, hadronization and the expanding background. Jet substructure, heavy flavor and jet--hadron correlations provide complementary constraints on these ingredients and can reveal information about QGP properties.

The comparison between hard and soft transport provides another test of the underlying dynamics. In a weakly coupled, quasiparticle medium dominated by small-angle scattering, kinetic estimates relate shear viscosity to the jet transport coefficient $\eta/s\sim T^3/\qhat_A$~\cite{MajumderViscosity}. The difference between the quark and gluon jet transport coefficients, $\qhat_A$ and $\qhat_F$, should be verified experimentally. Joint constraints can therefore test whether the same microscopic description accounts for bulk relaxation and the interaction of energetic probes with the QGP medium.

Future quantitative extraction of QGP properties requires inference that includes experimental correlations and theoretical uncertainties, consistency across collision energies and centralities, and further predictions for observables not used in the current calibration. Validated surrogate models and computational automation can extend these comparisons while preserving the physical constraints and numerical accuracy of the underlying calculations. Connecting jet modification to the medium's wake allows thermalization and the collective response of the QGP to be studied within a common dynamical description and constraints on the equation of state of the QGP.

{\bf Acknowledgment:} I want to thank all my collaborators on the work covered in this talk. This work is supported by the National Natural Science Foundation of China under Grant No. 12535010. AI assistance was used to refine the language of the original transcript of the talk.

\clearpage
\begingroup
\small
\setlength{\bibsep}{2pt}

\endgroup
\end{document}